\documentclass[12pt]{article}
\usepackage{amsmath}
\usepackage{bm}
\usepackage{amssymb}
\usepackage{amsthm}
\usepackage{longtable}
\usepackage{amsfonts}
\usepackage{geometry}
\usepackage{multirow}
\usepackage{multicol}
\usepackage{graphicx}
\usepackage{epsfig}
\usepackage{epstopdf}
\usepackage{subfigure}
\usepackage{color}
\usepackage{ulem}
\usepackage{booktabs}
\usepackage{rotating}
\usepackage[toc,page]{appendix}
\usepackage{cite}
\usepackage{cases}
\usepackage{makecell}
\begin{document}



\def\a{\alpha}
\def\b{\beta}
\def\d{\delta}
\def\e{\epsilon}
\def\g{\gamma}
\def\h{\mathfrak{h}}
\def\k{\kappa}
\def\l{\lambda}
\def\o{\omega}
\def\p{\wp}
\def\r{\rho}
\def\t{\tau}
\def\s{\sigma}
\def\z{\zeta}
\def\x{\xi}
\def\V={{{\bf\rm{V}}}}
 \def\A{{\cal{A}}}
 \def\B{{\cal{B}}}
 \def\C{{\cal{C}}}
 \def\D{{\cal{D}}}
\def\G{\Gamma}
\def\K{{\cal{K}}}
\def\O{\Omega}
\def\R{\bar{R}}
\def\T{{\cal{T}}}
\def\L{\Lambda}
\def\f{E_{\tau,\eta}(sl_2)}
\def\E{E_{\tau,\eta}(sl_n)}
\def\Zb{\mathbb{Z}}
\def\Cb{\mathbb{C}}

\def\R{\overline{R}}

\def\beq{\begin{equation}}
\def\eeq{\end{equation}}
\def\bea{\begin{eqnarray}}
\def\eea{\end{eqnarray}}
\def\ba{\begin{array}}
\def\ea{\end{array}}
\def\no{\nonumber}
\def\le{\langle}
\def\re{\rangle}
\def\lt{\left}
\def\rt{\right}

\baselineskip=20pt

\newfont{\elevenmib}{cmmib10 scaled\magstep1}
\newcommand{\preprint}{
   \begin{flushleft}
   \end{flushleft}\vspace{-1.3cm}
   \begin{flushright}\normalsize
   \end{flushright}}
\newcommand{\Title}[1]{{\baselineskip=26pt
   \begin{center} \Large \bf #1 \\ \ \\ \end{center}}}

\newcommand{\Author}{\begin{center}
	\large \bf
	Pengcheng Lu${\,}^{a}$,
    Junpeng Cao${\,}^{b,c,d,e}$,
    Wen-Li Yang${\,}^{e,f,g}$,
    Ian Marquette${\,}^{h}$
    and Yao-Zhong Zhang${\,}^{a}$
\end{center}}

\newcommand{\Address}{\begin{center}

    ${}^a$ School of Mathematics and Physics, The University of Queensland, Brisbane, QLD 4072, Australia\\
    ${}^b$ Beijing National Laboratory for Condensed Matter Physics, Institute of Physics, Chinese Academy of Sciences, Beijing 100190, China\\
	${}^c$ School of Physical Sciences, University of Chinese Academy of Sciences, Beijing 100049, China\\
	${}^d$ Songshan Lake Materials Laboratory, Dongguan, Guangdong 523808, China\\
	${}^e$ Peng Huanwu Center for Fundamental Theory, Xian 710127, China\\
    ${}^f$ Institute of Modern Physics, Northwest University, Xian 710127, China\\
    ${}^g$ Shaanxi Key Laboratory for Theoretical Physics Frontiers, Xian 710127, China\\
    ${}^h$ Department of Mathematical and Physical Sciences, La Trobe University, Bendigo, VIC 3552, Australia

\end{center}}

\Title{\Large Exact physical quantities of the $D_2^{(2)}$ spin chain model with generic open boundary conditions} \Author

\Address
\vspace{1cm}

\begin{abstract}

\noindent We study the quantum integrable spin chain model associated with the twisted $D_2^{(2)}$ algebra (or simply the $D_2^{(2)}$ model) under generic open boundary conditions. The Hamiltonian of this model can be factorized into the sum of two staggered XXZ spin chains. Applying the $t$-$W$ method, we derive the homogeneous Bethe ansatz equations for the zeros of the transfer matrix eigenvalues and the patterns of the corresponding zeros of the staggered XXZ spin chain with generic integrable boundaries. Based on these results, we analytically compute the surface energies and excitation energies of the $D_2^{(2)}$ model in different regimes of boundary parameters.

\vspace{1truecm} \noindent {\it PACS:}
75.10.Pq, 03.65.Vf, 71.10.Pm

\noindent {\it Keywords}: Spin chain; Reflection equation; Bethe
Ansatz; $T-Q$ relation
\end{abstract}
\newpage
\section{Introduction}
\label{intro} \setcounter{equation}{0}
The $D_2^{(2)}$ spin chain\cite{Jimbo1986537} is a quantum integrable model associated with the twisted quantum affine algebra $U_q(D^{(2)}_2)$ and has found applications in black holes and string theory. For instance, the relation of these models to the non-compact conformal field theory (CFT) and the $SL(2, R)/U(1)$ Euclidean black holes \cite{Witten1991314,
Maldacena20012929,Maldacena20012961,Hanany2002014} was first observed in \cite{Ikhlef2012081601}. Later this conjectured relation was revised and extended by Bazhanov et al. \cite{Bazhanov2012115337}. Besides, the authors in \cite{Robertson2021180} demonstrated that the $D_2^{(2)}$ spin chain with quantum group invariant boundary conditions is relevant to the lattice regularization of non-compact black hole boundary CFT, and this study has been further refined in \cite{Frahm2024149}. Also, the solution of this model provides the basis for solving models with higher-rank $D_n^{(2)}$ algebra symmetry using the nested Bethe ansatz method. During the past few decades, the $D_2^{(2)}$ model with periodic or diagonal boundary conditions has been extensively studied \cite{Reshetikhin1987235,Martins19997220,Martins2000721,Nepomechie2017924,Nepomechie2019266}.

Recently, the authors in \cite{Robertson2020144} reported an interesting result, indicating that the $R$-matrix of the $D_2^{(2)}$ model 
is closely related to the staggered XXZ spin chain and the antiferromagnetic Potts model\cite{Saleur1991219,Ikhlef2008483,Ikhlef2012081601}. Based on this, the exact solutions of the transfer matrices of both the closed and open $D_2^{(2)}$ chains with the help of factorization identities and the algebraic Bethe ansatz \cite{Nepomechie2021089}. Subsequently, the $D_2^{(2)}$ spin chain with generic non-diagonal boundary fields was solved in \cite{JHEP2022101} using the so-called off-diagonal Bethe ansatz. However, because of the non-homogeneity of the Bethe ansatz equations (BAEs) and the unclear distribution of the Bethe roots for the model, its physical properties in the thermodynamic limit remain an unsolved problem.

Based on characterizing the eigenvalues of the transfer matrix by their zeros instead of the traditional Bethe roots, a systematic $t$-$W$ method \cite{PRB1022020,PRBL1032021} has been proposed to compute the physical properties of integrable systems with or without $U(1)$-symmetry. In this method, homogeneous BAEs for the zeros of the transfer matrix eigenvalues can be obtained from the fusion relation of the transfer matrix. By applying the logarithm followed by differentiation to the homogeneous BAEs, the exact physical quantities of the systems in the thermodynamic limit can be calculated.
The $t$-$W$ method has been applied to many quantum integrable models associated with untwisted affine algebras, such as the XXZ spin chain\cite{PRBL1032021}, the spin-1 model \cite{JPA2024305202} and the $D_2^{(1)}$ spin chain \cite{NPB9842022} under generic open boundary conditions.

In our previous work \cite{Lu24arXiv}, we extended the $t$-$W$ method to the Izergin-Korepin model, which is a typical twisted $A$-type model.
In this paper, we apply the method to the spin chain model with twisted quantum affine algebra $U_q(D_2^{(2)})$ as underlying algebraic structure. Here, we consider the boundary condition to be the generic open one, thus the related underlying algebraic symmetry is broken. We derive the density of zeros, surface energy and excitation energy of the model using the $t$-$W$ method.

The paper is organized as follows.  Section 2 presents an introduction to the $D_2^{(2)}$ model with open boundaries and its integrability. The Hamiltonian of this model can be factorized into the sum of two staggered XXZ spin chains. In Section 3, we derive the exact solution of the system using the intrinsic properties of the transfer matrix and the zeros of its eigenvalues. The homogeneous BAEs for such zeros and the expression of the energy spectrum are given. In Section 4, we determine the detailed patterns of the zeros in the different regimes of the boundary parameters. Based on the patterns, the densities of the zeros and exact surface energies in all regimes are derived in Section 5. Section 6 is devoted to the computation of the bulk elementary excitations and the boundary excitations associated with the boundary fields. Concluding remarks are given in Section 7. 

\section{$D_2^{(2)}$ spin chain with generic open boundaries}
\label{Sec1} \setcounter{equation}{0}

Let $R(u)$ be the $R$-matrix associated with the 4-dimensional representation of the twisted quantum affine algebra $U_q(D_2^{(2)})$, where $u$ is the spectral parameter. The explicit matrix form of $R(u)$ can be found in \cite{Jimbo1986537,Martins2000721,Nepomechie2017924,Nepomechie2019266} and will be omitted here. By the standard procedure, the Hamiltonian of the $D_2^{(2)}$ spin chain model with generic boundary fields can be obtained from the transfer matrix $t(u)$ \cite{JPA198821}
\bea
H=\frac{\partial \ln t(u)}{ \partial u}\Big|_{u=0},\label{H-D22}
\eea
where $t(u)$ is defined by
\bea
t(u)=tr_0\{K_0^+(u)R_{01}(u)R_{02}(u)\cdots R_{0N}(u)K_0^-(u)R_{N0}(u)R_{N-10}(u)\cdots R_{10}(u)\}.\label{t-D22}
\eea
Here $tr_0$ means the partial trace in the
4-dimensional auxiliary space $V_0$, the subscript $j={1,\ldots,N}$ denotes the
4-dimensional  quantum or physical space $V_j$ at $j$-th site, and $K_0^{\pm}(u)$ are the reflection matrices given in \cite{Nepomechie201839LT02,Nepomechie2019434004,Malara2006P09013} in the auxiliary space at the left and right boundaries. 

Using the factorization method, the four-dimensional space $V_j$ can be regarded as the tensor product of the two-dimensional spaces, denoted by $V_{j^\prime}$. In the following, We will use the notations such as $V_1=V_{1^{\prime}}\otimes V_{2^{\prime}}$ and $V_2=V_{3^{\prime}}\otimes V_{4^{\prime}}$. Then the $R$-matrix of the $D_2^{(2)}$ spin chain can be factorized as the product of $R$-matrices of the anisotropic XXZ spin chain with suitable global transformation \cite{Robertson2020144,Nepomechie2021089}
\bea
&& R_{12}(u)=2^4[S\otimes S]\tilde
R_{1'4'}(u+i\pi)\tilde R_{1'3'}(u)
\tilde R_{2'4'}(u)\tilde R_{2'3'}(u-i\pi) [S\otimes S]^{-1},\label{Factor-R-1}\\[4pt]
&&R_{21}(u)=2^4[S\otimes S]\tilde R_{3'2'}(u+i\pi)\tilde R_{4'2'}(u)
\tilde R_{3'1'}(u)\tilde R_{4'1'}(u-i\pi) [S\otimes S]^{-1},\label{Factor-R-2}
\eea
where the transformation matrix $S$ is
\bea
 S=S^{-1}=\left(\begin{array}{cccc}
    1&&& \\
    &\frac{\cosh\frac \eta2}{\sqrt{\cosh\eta}}&-\frac{\sinh\frac \eta2}{\sqrt{\cosh\eta}}& \\[4pt]
    &-\frac{\sinh\frac \eta2}{\sqrt{\cosh\eta}}&-\frac{\cosh\frac \eta2}{\sqrt{\cosh\eta}}& \\[4pt]
    &&&1\end{array}\right), \label{s1-matrix}
    \eea
and the $R$-matrix $\tilde R(u)$ in the tensor space $V_{1'}\otimes V_{2'}$ reads
\bea
 \tilde R_{1'2'}(u)=\left(\begin{array}{cccc}
    \sinh(-\frac u2+\eta)&&&\\
    &\sinh \frac u2 &e^{-\frac u2}\sinh \eta &  \\[4pt]
    &e^{\frac u2}\sinh \eta &\sinh\frac u2 &  \\[4pt]
    &&&\sinh(-\frac u2+\eta)
    \end{array}\right).\label{Rs-matrix}
\eea
It satisfies the quantum Yang-Baxter equation (QYBE)
\bea
\tilde R_{1'2'}(u-v)\tilde R_{1'3'}(u)\tilde R_{2'3'}(v)
=\tilde R_{2'3'}(v)\tilde R_{1'3'}(u)\tilde R_{1'2'}(u-v),\label{QYB}\eea
and the following properties
\bea
&&\hspace{-1.5cm}\mbox{ Initial
condition}:\hspace{0.3cm}\,\tilde R_{1'2'}(0)= \sinh\eta \tilde P_{1'2'},\label{Int-R}\\
&&\hspace{-1.5cm}\mbox{ PT-symmetry}:\hspace{0.7cm}\,\tilde R_{2'1'}(u)=\tilde R^{t_{1'}\,t_{2'}}_{1'2'}(u),\label{PT}\\
&&\hspace{-1.5cm}\mbox{ Quasi--period}: \,
\qquad \tilde R_{1'2'}(u+2i\pi)=-\tilde R_{1'2'}(u),\label{Periodic}\\
&&\hspace{-1.5cm}\mbox{ Unitarity
relation}:\,\tilde R_{1'2'}(u)\tilde R_{2'1'}(-u)= \rho_s(u)=\,\sinh(-\frac{u}{2}+\eta)\sinh(\frac{u}{2}+\eta),\label{Unitarity}\\
&&\hspace{-1.5cm}\mbox{ Crossing
relation}:\,\,\tilde R^{t_{1'}}_{1'2'}(u)\tilde M_{1'}\tilde R^{t_{1'}}_{2'1'}(u)\tilde M^{-1}_{1'}=\,\rho_s(u-\eta),
\label{crosing}
\eea

In similar fashion, the reflection matrices $K_{1}^{\pm}(u)$ of the $D_2^{(2)}$ spin chain have the following factorized form
\bea
\hspace{-1truecm}&&K_1^+(u)=[\rho_s(i\pi)]^{-\frac{1}{2}}S\tilde R_{2'1'}(i\pi)\tilde K_{2'}^+(u) \tilde M_{2'}^{-1} \tilde R_{1'2'}(-2u+4\eta-i\pi) \tilde M_{2'}\tilde K_{1'}^+(u+i\pi)S^{-1},\label{Factor-K-1}\\
\hspace{-1truecm}&&K_1^-(u)=[\rho_s(i\pi)]^{-\frac{1}{2}}S
\tilde K_{1'}^-(u+i\pi)\tilde R_{2'1'}(2u+i\pi)\tilde K_{2'}^-(u)\tilde R_{1'2'}(-i\pi)S^{-1},\label{Factor-K-2}
\eea
where $\tilde M_{k'}$ is the diagonal constant matrix $\tilde M_{k'}=diag(e^{\eta},e^{-\eta})$, and $\tilde K_{k'}^{\pm}(u)$ are the $2\times 2$ generic non-diagonal reflection matrices of the XXZ spin chain \cite{Vega19946129}
\bea
\tilde K_{k'}^{-}(u)&=&\left(
  \begin{array}{cccc}
    -e^{-\frac{u}{2}}\sinh(\frac{u}{2}-s) &  s_1\sinh u\\
    s_2\sinh u  & e^{\frac{u}{2}}\sinh(\frac{u}{2}+s)\\
  \end{array}
\right),\label{Ksn}\\
\tilde K_{k'}^{+}(u)&=&\tilde M_{k'}\tilde K_{k'}^{-}(-u+2\eta)|_{(s,s_1,s_2)\rightarrow(s',s_1',s_2')}\label{Ksp},
\eea
which satisfy the reflection equations
\bea
\hspace{-1truecm}&&\tilde R_{1'2'}(u-v)\tilde K_{1'}^{-}(u)\tilde R_{2'1'}(u+v)\tilde K_{2'}^{-}(v)=\tilde K_{2'}^{-}(v)\tilde R_{2'1'}(u+v)\tilde K_{1'}^{-}(u)\tilde R_{1'2'}(u-v),\label{REs1}\\
\hspace{-1truecm}&&\tilde R_{1'2'}(-u+v)\tilde K_{1'}^{+}(u)M_{1'}^{-1}\tilde R_{2'1'}(-u-v+4\eta)M_{1'}\tilde K_{2'}^{-}(v)=\no\\
\hspace{-1truecm}&&\hspace{4truecm}\tilde K_{2'}^{-}(v)M_{1'}\tilde R_{2'1'}(-u-v+4\eta)M_{1'}^{-1}\tilde K_{1'}^{+}(u)\tilde R_{1'2'}(u-v).\label{REs2}
\eea
The six free parameters $\{s, s_1, s_2\}$ and $\{s', s_1', s_2'\}$ in the matrices $\tilde K_{k'}^{\pm}(u)$ describe the left and right boundary fields, respectively.

From the $R$-matrix $\tilde R(u)$ (\ref{Rs-matrix}), we can define the one-row monodromy matrices $\tilde T_{0'}(u)$ and $\hat{ \tilde T}_{0'}(u)$, which are two $2\times 2$ matrices with operator-valued elements acting on quantum space $V^{\otimes {(2N)}'}$,
\bea
&&\tilde T_{0'}(u)=\tilde R_{0'1'}(u-\theta_1)\tilde R_{0'2'}(u-\theta_2)\cdots \tilde R_{0'(2N)'}(u-\theta_{2N}),\\
&&\hat{ \tilde T}_{0'}(u)=\tilde R_{0'(2N)'}(u+\theta_{2N})\tilde R_{0'(2N-1)'}(u+\theta_{2N-1})\cdots \tilde R_{0'1'}(u-\theta_{1}),
\eea
where $V_{0'}$ is a $2\times 2$ auxiliary space and $\{\theta_j|j=1,\cdots,2N\}$ are the inhomogeneous parameters. We can now construct the transfer matrix of the inhomogeneous XXZ spin chain as
\bea
\tilde t(u)=tr_{0'}\{\tilde K_{0'}^{+}(u)\tilde T_{0'}(u)\tilde K_{0'}^{-}(u)\hat{ \tilde T}_{0'}(u) \},\label{ts}
\eea
where $tr_{0'}$ means the partial trace over the auxiliary space $V_{0'}$.
Based on the factorized forms (\ref{Factor-R-1}), (\ref{Factor-R-2}), (\ref{Factor-K-1}) and (\ref{Factor-K-2}), the transfer matrix $t(u)$ (\ref{t-D22}) can be expressed as the product of transfer matrices of the two staggered XXZ spin chains with a fixed spectral difference \cite{JHEP2022101}
\bea\label{Factor-t}
t(u)=2^{8N}\rho_s(2u+i\pi-2\eta)\tilde{t}_s(u+i\pi)\tilde{t}_s(u),
\eea
where $\tilde{t}_s(u)=\tilde{t}(u)|_{\{\theta_j\}=\{0,i\pi\}}$ and $\{\theta_j\}=\{0,i\pi\}$ means that the inhomogeneous parameters are staggered, i.e., $\theta_j=0$ for the odd $j$ and $\theta_j=i\pi$ for the even $j$. Therefore, the Hamiltonian (\ref{H-D22}) is given by the sum of two staggered XXZ spin chains
\bea
H=\hspace{-0.6truecm}&&\frac{\partial \ln \tilde{t}_s(u)}{ \partial u}\Big|_{u=0}+\frac{\partial \ln \tilde{t}_s(u+i\pi)}{ \partial u}\Big|_{u=0}-\tanh(2\eta)\no\\[6pt]
=\hspace{-0.6truecm}&&H_1+H_2-\tanh(2\eta),\label{Hs}
\eea
where the Hamiltonians $H_1$ and $H_2$ can be expressed in terms of the Pauli matrices as
\bea
H_1=\hspace{-0.6truecm}&&\frac{\partial \ln \tilde{t}_s(u)}{ \partial u}\Big|_{u=0}\no\\[6pt]
=\hspace{-0.6truecm}&&-\sum_{j=1}^{N-1} \Big[\frac{\sigma^z_{{(2j-1)}^{\prime}} }{2} -\frac{\sigma^z_{{(2j+1)}^{\prime}} }{2}+\frac{\tanh(\eta)}{2}\sigma^z_{{(2j-1)}^{\prime}}\sigma^z_{{(2j)}^{\prime}}+\mathrm{csch}(2\eta)\sigma^z_{{(2j-1)}^{\prime}}\sigma^z_{{(2j+1)}^{\prime}}   \no\\[6pt]
\hspace{-0.6truecm}&&+\frac{\tanh(\eta)}{2}\sigma^z_{{(2j)}^{\prime}}\sigma^z_{{(2j+1)}^{\prime}}+\mathrm{csch}(2\eta)(\sigma^x_{{(2j-1)}^{\prime}}\sigma^x_{{(2j+1)}^{\prime}}+\sigma^y_{{(2j-1)}^{\prime}}\sigma^y_{{(2j+1)}^{\prime}})\no\\[6pt]
\hspace{-0.6truecm}&&+\frac{\mathrm{sech} (\eta)}{2}\Big((\sigma^x_{{(2j-1)}^{\prime}}\sigma^x_{{(2j)}^{\prime}}\hspace{-0.08truecm}+\hspace{-0.08truecm}\sigma^y_{{(2j-1)}^{\prime}}\sigma^y_{{(2j)}^{\prime}})\sigma^z_{{(2j+1)}^{\prime}}  \hspace{-0.08truecm}+\hspace{-0.08truecm}\sigma^z_{{(2j-1)}^{\prime}}(\sigma^x_{{(2j)}^{\prime}}\sigma^x_{{(2j+1)}^{\prime}}\hspace{-0.08truecm}+\hspace{-0.08truecm}\sigma^y_{{(2j)}^{\prime}}\sigma^y_{{(2j+1)}^{\prime}})\Big)\Big]\no\\[6pt]
\hspace{-0.6truecm}&&-\hspace{-0.08truecm}\left(\frac{1}{2}\hspace{-0.08truecm}+\hspace{-0.08truecm}\frac{1}{2}\coth(s)\mathrm{sech}^2(\eta)\right)\sigma^z_{{(2N-1)}^{\prime}}\hspace{-0.08truecm}-\hspace{-0.08truecm}\frac{1}{2}\coth(s)\tanh^2(\eta)\sigma^z_{{(2N)}^{\prime}}\hspace{-0.08truecm}-\hspace{-0.08truecm}\frac{\tanh(\eta)}{2}\sigma^z_{{(2N-1)}^{\prime}}\sigma^z_{{(2N)}^{\prime}}\no\\[6pt]
\hspace{-0.6truecm}&&+\tanh(\eta)\mathrm{csch}(s)\Big(s_1\sigma^z_{{(2N-1)}^{\prime}}\sigma^+_{{(2N)}^{\prime}}\hspace{-0.08truecm}+\hspace{-0.08truecm}s_2 \sigma^z_{{(2N-1)}^{\prime}}\sigma^-_{{(2N)}^{\prime}}\Big)\hspace{-0.08truecm}-\hspace{-0.08truecm}s_1\mathrm{csch}(\eta)\mathrm{sech}(\eta)\sigma^+_{{(2N-1)}^{\prime}}\no\\[6pt]
\hspace{-0.6truecm}&&-\frac{1}{2}\tanh(\eta)\coth(\eta)\mathrm{sech}(\eta)\Big(\sigma^x_{{(2N-1)}^{\prime}}\sigma^x_{{(2N)}^{\prime}}\hspace{-0.08truecm}+\hspace{-0.08truecm}\sigma^y_{{(2N-1)}^{\prime}}\sigma^y_{{(2N)}^{\prime}} \Big)\hspace{-0.08truecm}-\hspace{-0.08truecm}s_2\mathrm{csch}(\eta)\mathrm{sech}(\eta)\sigma^-_{{(2N-1)}^{\prime}}\no\\[6pt]
\hspace{-0.6truecm}&&+\mathrm{csch}(s^{\prime})\Big(\frac{e^{s^{\prime}}\sigma^z_{{1}^{\prime}}}{2}+e^{\eta}{s_1}^{\prime}\sigma^+_{{1}^{\prime}}+e^{-\eta}{s_2}^{\prime} \sigma^-_{{1}^{\prime}}  \Big)-N\coth(2\eta)-\frac{\tanh(\eta)}{2},
\eea
\bea
H_2=\hspace{-0.6truecm}&&\frac{\partial \ln \tilde{t}_s(u+i\pi)}{ \partial u}\Big|_{u=0}\no\\[6pt]
=\hspace{-0.6truecm}&&-\hspace{-0.06truecm}\sum_{j=1}^{N-1} \Big[\frac{\sigma^z_{{(2j)}^{\prime}} }{2} -\frac{\sigma^z_{{(2j+2)}^{\prime}} }{2}+\frac{\tanh(\eta)}{2}\sigma^z_{{(2j)}^{\prime}}\sigma^z_{{(2j+1)}^{\prime}}+\mathrm{csch}(2\eta)\sigma^z_{{(2j)}^{\prime}}\sigma^z_{{(2j+2)}^{\prime}}   \no\\[6pt]
\hspace{-0.6truecm}&&+\frac{\tanh(\eta)}{2}\sigma^z_{{(2j+1)}^{\prime}}\sigma^z_{{(2j+2)}^{\prime}}+\mathrm{csch}(2\eta)(\sigma^x_{{(2j)}^{\prime}}\sigma^x_{{(2j+2)}^{\prime}}+\sigma^y_{{(2j)}^{\prime}}\sigma^y_{{(2j+2)}^{\prime}})\no\\[6pt]
\hspace{-0.6truecm}&&+\frac{\mathrm{sech} (\eta)}{2}\Big((\sigma^x_{{(2j)}^{\prime}}\sigma^x_{{(2j+1)}^{\prime}}\hspace{-0.08truecm}+\hspace{-0.08truecm}\sigma^y_{{(2j)}^{\prime}}\sigma^y_{{(2j+1)}^{\prime}})\sigma^z_{{(2j+2)}^{\prime}}  \hspace{-0.08truecm}+\hspace{-0.08truecm}\sigma^z_{{(2j)}^{\prime}}(\sigma^x_{{(2j+1)}^{\prime}}\sigma^x_{{(2j+2)}^{\prime}}\hspace{-0.08truecm}+\hspace{-0.08truecm}\sigma^y_{{(2j+1)}^{\prime}}\sigma^y_{{(2j+2)}^{\prime}})\Big)\Big]\no\\[6pt]
\hspace{-0.6truecm}&&-\frac{1}{2}\mathrm{sech}^2(\eta)\mathrm{sech}(s^{\prime})\Big[\sinh(\eta)\sinh(s^{\prime})(\sigma^x_{{1}^{\prime}}\sigma^x_{{2}^{\prime}}+\sigma^y_{{1}^{\prime}}\sigma^y_{{2}^{\prime}})    -\sinh^2(\eta)\sinh(s^{\prime})\sigma^z_{{1}^{\prime}}\no\\[6pt]
\hspace{-0.6truecm}&&-(\cosh^2(\eta)\cosh(s^{\prime})+\sinh(s^{\prime})) \sigma^z_{{2}^{\prime}}+\frac{\sinh(2\eta)}{2}\cosh(s^{\prime})\sigma^z_{{1}^{\prime}}\sigma^z_{{2}^{\prime}}+2e^{\eta}s_1^{\prime}\cosh(\eta)\sigma^+_{{2}^{\prime}}\no\\[6pt]
\hspace{-0.6truecm}&&+2e^{-\eta}s_2^{\prime}\cosh(\eta)\sigma^-_{{2}^{\prime}}+e^{\eta}s_1^{\prime}\sinh(2\eta)\sigma^+_{{1}^{\prime}}\sigma^z_{{2}^{\prime}}+e^{-\eta}s_2^{\prime}\sinh(2\eta)\sigma^-_{{1}^{\prime}}\sigma^z_{{2}^{\prime}}  \Big]\hspace{-0.08truecm}-\hspace{-0.08truecm}(N\hspace{-0.08truecm}-\hspace{-0.08truecm}1)\coth(2\eta)\no\\[6pt]
\hspace{-0.6truecm}&&-\mathrm{sech}(s)\Big(\frac{e^{s}\sigma^z_{{(2N)}^{\prime}}}{2}\hspace{-0.08truecm}-\hspace{-0.08truecm}s_1\sigma^+_{{(2N)}^{\prime}}\hspace{-0.08truecm}-\hspace{-0.08truecm}s_2\sigma^-_{{(2N)}^{\prime}}\Big) 
\hspace{-0.08truecm}-\hspace{-0.08truecm}\frac{ (2+\coth^2(\eta))\tanh(\eta)}{2}.
\eea
From the QYBE (\ref{QYB}), the reflection equations (\ref{REs1})-(\ref{REs2}) and factorization (\ref{Factor-t}), the commutation relations $[\tilde t(u),\tilde t(v)]=0$ and $[t(u),t(v)]=0$ \cite{JPA198821} can be easily derived, which ensure the integrability of the Hamiltonian (\ref{Hs}).

Some remarks are in order. In terms of spin matrices, the Hamiltonian (2.23) of the $D_2^{(2)}$ spin chain model contains next-nearest-neighbour interactions, which result from the chosen inhomogeneities in the staggered XXZ spin chain. The fusion method requires a particular choice of this staggering and is used to construct higher spin models by projecting onto a subset of states.

\section{Exact solution}
\label{Exact solution} \setcounter{equation}{0}
Using the properties of the $R$-matrices $\tilde{R}(u)$ (\ref{Rs-matrix}) and the $K$-matrices $\tilde{K}^{\pm}(u)$ (\ref{Ksn})-(\ref{Ksp}), one can prove that the transfer matrix $\tilde t(u)$ (\ref{ts}) satisfies the following operator identities and asymptotic
behaviour \cite{Faldella2014P01011,ODBA2013}
\bea
&&\hspace{-1truecm}\tilde{t}(\pm \theta_j)\tilde{t}(\pm \theta_j+2\eta)=\frac{4\sinh(\pm \theta_j-2\eta)\sinh(\pm \theta_j+2\eta)}{\alpha\alpha^{\prime}\sinh(\pm \theta_j-\eta)\sinh(\pm \theta_j+\eta)}\cosh\frac{\pm\theta_j-\alpha_1}{2}\no\\[8pt]
&&\times \cosh\frac{\pm \theta_j+\alpha_1}{2}\cosh\frac{\pm \theta_j-\alpha_2}{2}\cosh\frac{\pm \theta_j+\alpha_2}{2}\cosh\frac{\pm \theta_j-\alpha_1^{\prime}}{2}\cosh\frac{\pm \theta_j+\alpha_1^{\prime}}{2}\no\\[8pt]
&&\times\cosh\frac{\pm \theta_j-\alpha_2^{\prime}}{2}\cosh\frac{\pm \theta_j+\alpha_2^{\prime}}{2}\prod_{l=1}^{2N}\sinh\frac{\pm \theta_j-\theta_l-2\eta}{2}\sinh\frac{\pm \theta_j-\theta_l+2\eta}{2}\no\\[8pt]
&&\times \sinh\frac{\pm \theta_j+\theta_l-2\eta}{2}\sinh\frac{\pm \theta_j+\theta_l+2\eta}{2},\quad j=1,\cdots,2N,\\[8pt]
&&\hspace{-1truecm}\tilde{t}(u)|_{u\rightarrow \pm \infty}=-2^{-4N-2}e^{\mp[(2N+2)\eta]}(e^{-\eta}s_1s_2^{\prime}+e^{\eta}s_2s_1^{\prime})\times {\rm id }+\cdots,
\eea
where the related constants are defined as
\bea
&&\alpha=\frac{1}{2s_1s_2},\quad \beta=\sqrt{\frac{8s_1s_2\cosh(2s)+16(s_1s_2)^2+1 }{16(s_1s_2)^2} }, \cosh\alpha_1=\frac{\alpha}{2}+\beta,\no\\[8pt]
&&\cosh\alpha_2=\frac{\alpha}{2}-\beta,\quad \alpha^{\prime}=\frac{1}{2s_1^{\prime}s_2^{\prime}},\quad \beta^{\prime}=\sqrt{\frac{8s_1^{\prime}s_2^{\prime}\cosh(2s^{\prime})+16(s_1^{\prime}s_2^{\prime})^2+1 }{16(s_1^{\prime}s_2^{\prime})^2} },\no\\[8pt]
&&\cosh\alpha_1^\prime=\frac{\alpha^\prime}{2}+\beta^\prime,\quad \cosh\alpha_2^\prime=\frac{\alpha^\prime}{2}-\beta^\prime.\label{aa1a2}
\eea
In addition, the values of $\tilde t(u)$ at the points of $u=0,\,2\eta,\,i\pi$ can be calculated directly
\bea
&&\hspace{-1truecm}\tilde{t}(0)=\tilde{t}(2\eta)=2\cosh\eta\sinh s\sinh s^{\prime}\prod_{j=1}^{2N}\rho_s(\theta_j),\no\\[8pt]
&&\hspace{-1truecm}\tilde{t}(i\pi)=2\cosh \eta \cosh s \cosh s^{\prime}\prod_{j=1}^{2N}\rho_s(\theta_j+i\pi),
\eea
where the function $\rho_s(u)$ is given by (\ref{Unitarity}).

In this paper, we will focus on the case of the anisotropic parameter $\eta>0$. The hermiticity of the Hamiltonian (\ref{H-D22}) requires that $s$ and $s^{\prime}$ are real, $s_2=s_1^{*}$ and $s_2^{\prime}=s_1^{\prime*}\,e^{2\eta}$, where the superscript $*$ means that the complex conjugate.
According to the parametrization (\ref{aa1a2}), the hermiticity requires that the boundary parameters
$\alpha_1$ and $\alpha_1^{\prime}$ are real, $\alpha_2={\rm Re}(\alpha_2)+i\pi$ and $\alpha_2^{\prime}={\rm Re}(\alpha_2^{\prime})+i\pi$.

Let $\tilde\Lambda(u)$ denote an eigenvalue of the transfer matrix $\tilde t(u)$. From the results on $\tilde{t}(u)$ presented above, it follows that the eigenvalue $\tilde{\Lambda}(u)$ satisfies
\bea
&&\hspace{-1truecm}\tilde{\Lambda}(0)=\tilde{\Lambda}(2\eta)=2\cosh\eta\sinh s\sinh s^{\prime}\prod_{j=1}^{2N}\rho_s(\theta_j),\label{sp1}\\[8pt]
&&\hspace{-1truecm}\tilde{\Lambda}(i\pi)=2\cosh \eta \cosh s \cosh s^{\prime}\prod_{j=1}^{2N}\rho_s(\theta_j+i\pi),\label{sp2}\\[8pt]
&&\hspace{-1truecm}\tilde{\Lambda}(u)|_{u\rightarrow \pm \infty}=-2^{-4N-2}e^{\pm[(2N+2)(u-\eta)]}(e^{-\eta}s_1s_2^{\prime}+e^{\eta}s_2s_1^{\prime}),\label{ab}\\[8pt]
&&\hspace{-1truecm}\tilde{\Lambda}(\pm \theta_j)\tilde{\Lambda}(\pm \theta_j+2\eta)=\frac{4\sinh(\pm \theta_j-2\eta)\sinh(\pm \theta_j+2\eta)}{\alpha\alpha^{\prime}\sinh(\pm \theta_j-\eta)\sinh(\pm \theta_j+\eta)}\no\\[8pt]
&&\times\cosh\frac{\pm\theta_j-\alpha_1}{2} \cosh\frac{\pm \theta_j+\alpha_1}{2}\cosh\frac{\pm \theta_j-\alpha_2}{2}\cosh\frac{\pm \theta_j+\alpha_2}{2}\no\\[8pt]
&&\times\cosh\frac{\pm \theta_j-\alpha_1^{\prime}}{2}\cosh\frac{\pm \theta_j+\alpha_1^{\prime}}{2}\cosh\frac{\pm \theta_j-\alpha_2^{\prime}}{2}\cosh\frac{\pm \theta_j+\alpha_2^{\prime}}{2}\no\\[8pt]
&&\times \prod_{l=1}^{2N}\sinh\frac{\pm \theta_j-\theta_l-2\eta}{2}\sinh\frac{\pm \theta_j-\theta_l+2\eta}{2}\sinh\frac{\pm \theta_j+\theta_l-2\eta}{2}\sinh\frac{\pm \theta_j+\theta_l+2\eta}{2},\no\\
  \label{id-lambda}
\eea
where $j=1,\cdots,2N$. From the above relations and the definition (\ref{ts}) of the transfer matrix $\tilde{t}(u)$, we know that $\tilde{\Lambda}(u)$ is a degree $4N+4$ polynomial of $\displaystyle e^{u/2}$ and can be parameterized in terms of its zeros as
\bea
\tilde{\Lambda}(u)=\tilde{\Lambda}_0\prod_{j=1}^{2N+2}\sinh(\frac{u}{2}-\frac{z_j}{2}-\frac{\eta}{2})\sinh(\frac{u}{2}+\frac{z_j}{2}-\frac{\eta}{2}),\label{Lamz}
\eea
where $\tilde{\Lambda}_0$ is a coefficient and $\{z_j|j=1,\cdots,2N+2\}$ are the zeros of the polynomial. Putting the parameterization (\ref{Lamz}) into (\ref{id-lambda}), we can obtain the homogeneous BAEs
\bea
&&\tilde{\Lambda}_0^2\prod_{k=1}^{2N+2}\sinh(\frac{\theta_j}{2}-\frac{z_k}{2}-\frac{\eta}{2})\sinh(\frac{\theta_j}{2}+\frac{z_k}{2}-\frac{\eta}{2})\sinh(\frac{\theta_j}{2}-\frac{z_k}{2}+\frac{\eta}{2})\sinh(\frac{\theta_j}{2}+\frac{z_k}{2}+\frac{\eta}{2})\no\\
&&\qquad\quad =\frac{4\sinh(\theta_j-2\eta)\sinh( \theta_j+2\eta)}{\alpha\alpha^{\prime}\sinh( \theta_j-\eta)\sinh(\theta_j+\eta)}\cosh\frac{\theta_j-\alpha_1}{2}\cosh\frac{ \theta_j+\alpha_1}{2}\cosh\frac{\theta_j-\alpha_2}{2}\no\\[8pt]
&&\qquad\qquad \times \cosh\frac{\theta_j+\alpha_2}{2}\cosh\frac{\theta_j-\alpha_1^{\prime}}{2}\cosh\frac{\theta_j+\alpha_1^{\prime}}{2}\cosh\frac{\theta_j-\alpha_2^{\prime}}{2}\cosh\frac{\theta_j+\alpha_2^{\prime}}{2}\no\\[8pt]
&&\qquad\qquad\times\prod_{l=1}^{2N}\sinh\frac{\theta_j-\theta_l-2\eta}{2}\sinh\frac{\theta_j-\theta_l+2\eta}{2} \sinh\frac{\theta_j+\theta_l-2\eta}{2}\sinh\frac{\theta_j+\theta_l+2\eta}{2},\no\\
\label{ba-zeros}
\eea
where $j=1,\cdots,N$. The BAEs and Eqs.(\ref{sp1})-(\ref{ab}) completely determine the $2N+3$ unknowns $\tilde{\Lambda}_0$ and $\{z_j|j=1,\cdots,2N+2\}$. Moreover, the energy spectrum of the Hamiltonian (\ref{H-D22}) can be determined by the zeros as
\bea\label{Energy-expressing}
E=-2\pi\sum_{j=1}^{2N+2} [a_1(i z_j)+a_1(i z_j+\pi)]-\tanh(2\eta),
\eea
where $$ a_n(u)=\frac{1}{2\pi i}\partial_u\left[\ln\sin(\frac{u}{2}-\frac{in\eta}{2})-\ln\sin(\frac{u}{2}+\frac{in\eta}{2})\right]=\frac{1}{2\pi}\frac{\sinh(n\eta)}{\cosh(n\eta)-\cos u}.$$

\section{Zero patterns}
\label{Patterns} \setcounter{equation}{0}

We first study the solutions of zeros $\{z_j\}$ for the ground state. Without loss of generality, we let $\{z_j=-i \bar z_j\}$, and choose the inhomogeneity parameters as $\{ \theta_{2l-1}=i\bar \theta_{2l-1},\theta_{2l}=i\bar \theta_{2l}+i\pi|l=1,\ldots,N\}$ with real $\bar \theta_l$. For notational simplicity, we set $\bar{\alpha}_2={\rm Re} ({\alpha}_2)$ and $\bar{\alpha}^{\prime}_2={\rm Re} ({\alpha}^{\prime}_2)$.

\begin{figure}[htbp]
\centering
\includegraphics[scale=0.6]{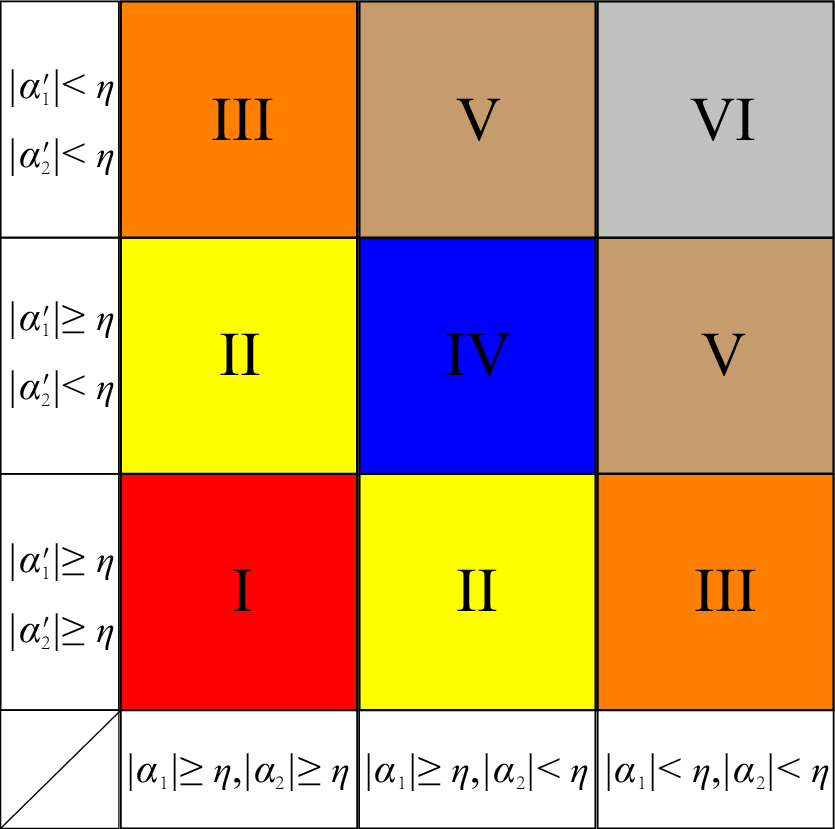}
  \caption{The distribution of $\bar{z}$-zeros for the ground state according to the boundary parameters.}\label{fig-region}
\end{figure}

By systematically varying the boundary parameters across their respective ranges with a fixed step size, we obtain the solutions of Eq.(\ref{ba-zeros}) with $N=4$ and $\eta=0.35$.
Summarizing the distributions of solutions and carrying out the singularity analysis in the thermodynamic limit \cite{le}, we find that the patterns of the $\bar{z}$-zeros include: (1) a set of real zeros $\{\tilde{z}_j\}$; (2) boundary pairs either at the origin or at ${\rm Re}\{\bar{z}_j\}=\pi$, where the imaginary parts of conjugate pairs are determined by the boundary parameters; (3) two extra conjugate pairs at the origin and at ${\rm Re}\{\bar{z}_j\}=\pi$, respectively. From the number and type of the boundary pairs, the distribution of the $\bar{z}$ zeros can be divided into six different regimes, as shown in Fig.\ref{fig-region}.

\begin{figure}[htbp]
  \centering
\subfigure{\label{fig21:subfig:a}     \includegraphics[scale=0.62]{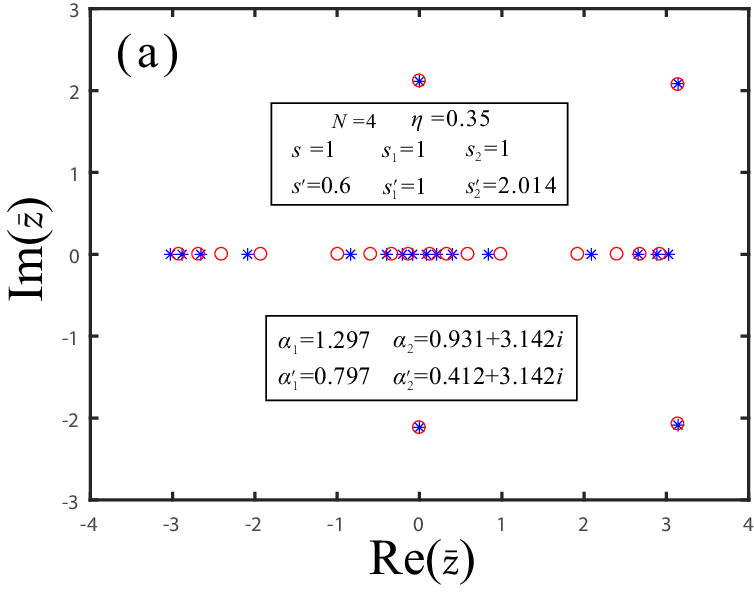}
    }  \subfigure{\label{fig21:subfig:b}   \includegraphics[scale=0.62]{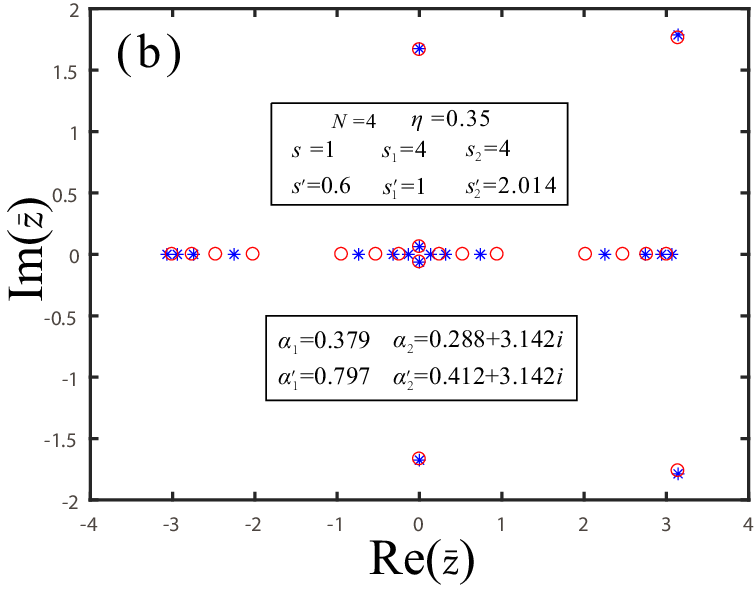}
    }
  \caption{Pattern of $\bar{z}$-zeros  with certain model parameters for the ground state. (a) The boundary parameters are chosen in regime I. (b) The boundary parameters are chosen in regime II. The blue asterisks indicate $\bar{z}$-zeros for $\{\bar{\theta}_j=0|j=1,\cdots,2N\}$ and the red circles specify $\bar{z}$-zeros
   with the inhomogeneity parameters $\{\bar{\theta}_j=0.1 j|j=1,\cdots,2N\}$.}\label{fig41}
\end{figure}

1) In regime I, where $\{|\alpha_1|,|\bar{\alpha}_2|, |\alpha_1^{'}|,|\bar{\alpha}_2^{\prime}|\}\geq \eta$, as shown in Fig.\ref{fig21:subfig:a}, all $\bar{z}$ zeros form $4N$ real zeros and two extra pairs $\pm i\chi_1$, $\pi\pm i\chi_2$ with real $\chi_1$, $\chi_2$. In the thermodynamic limit, $\chi_1$ and $\chi_2$ would tend to infinity.

2) In regime II, where $\{|\alpha_1|, |\alpha_1^{\prime}|,|\bar{\alpha}_2^{\prime}|\}\geq \eta$, $|\bar{\alpha}_2|< \eta$ or $\{|\alpha_1|, |\alpha_1^{\prime}|,|\bar{\alpha}_2|\}\geq \eta$,  $|\bar{\alpha}^{\prime}_2|< \eta$
, as shown in Fig.\ref{fig21:subfig:b}, all the $\bar{z}$-zeros form $4N-2$ real zeros, two extra pairs and one boundary conjugate pair $\pm i(\eta-{\{|\bar{\alpha}_2|,|\bar{\alpha}^{\prime}_2|\}}_{\rm min})$.

\begin{figure}[htbp]
  \centering
\subfigure{\label{fig32:subfig:a} \includegraphics[scale=0.62]{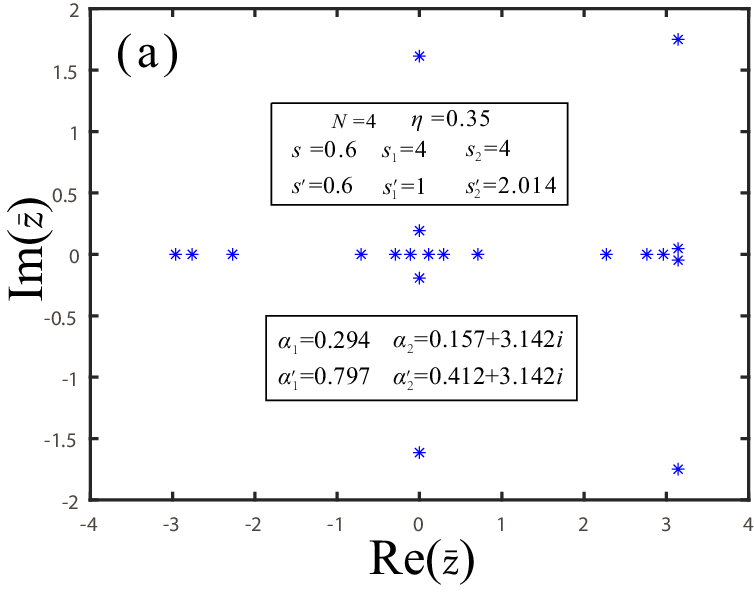} }\subfigure{\label{fig32:subfig:b} \includegraphics[scale=0.62]{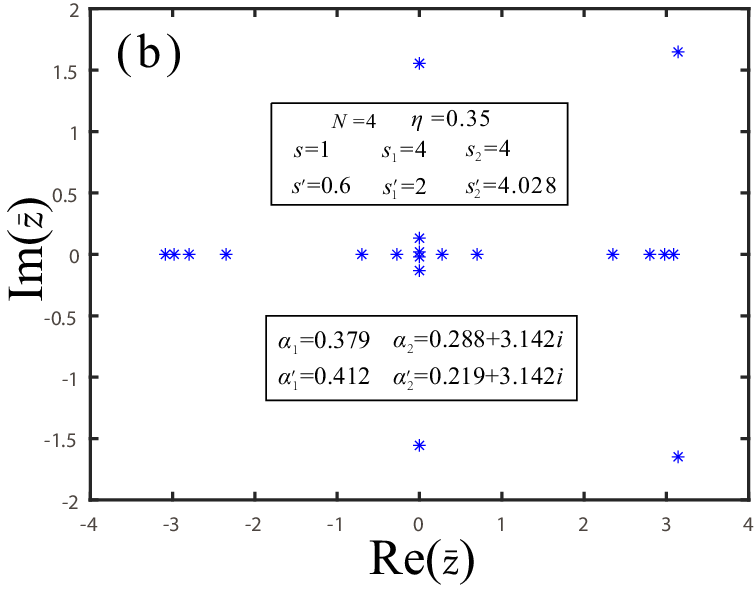}    } \subfigure{\label{fig32:subfig:c}\includegraphics[scale=0.62]{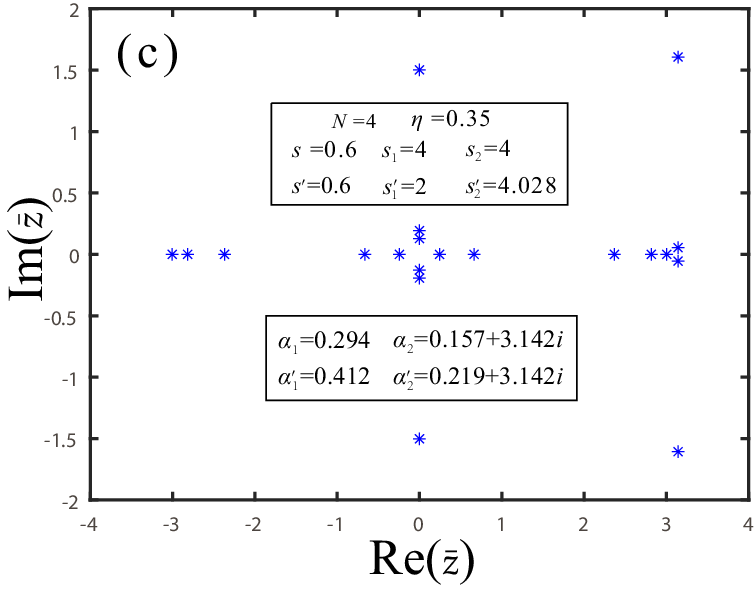}}
\subfigure{\label{fig32:subfig:d} \includegraphics[scale=0.62]{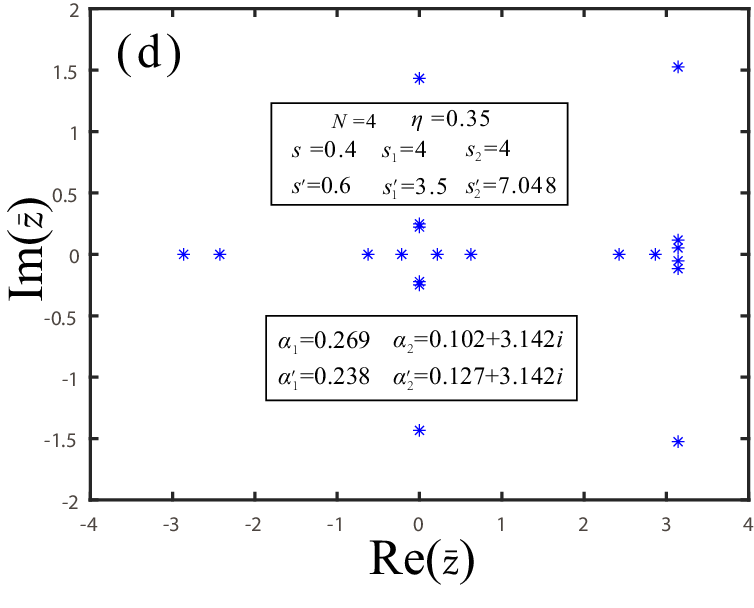}}
  \caption{Patterns of $\bar{z}$-zeros with certain model parameters for $\{\bar{\theta}_j= 0|j=1,\cdots,2N\}$ at the ground state. The boundary parameters in $(a)$-$(d)$ are chosen in the regimes III-VI, respectively.}\label{fig42}
\end{figure}

3) In regime III, where $\{|\alpha_1|,|\bar{\alpha}_2|\}\geq \eta$, $\{|\alpha^{\prime}_1|,|\bar{\alpha}^{\prime}_2|\}< \eta$ or $\{|\alpha_1|,|\bar{\alpha}_2|\}< \eta$, $\{|\alpha^{\prime}_1|,|\bar{\alpha}^{\prime}_2|\}\geq \eta$, as shown in Fig.\ref{fig32:subfig:a}, all $\bar{z}$-zeros form $4N-2$ real zeros, two extra pairs and two boundary conjugate pairs $\pi\pm i(\eta-{\{|{\alpha}_1|,|{\alpha}^{\prime}_1|\}}_{\rm min})$, $\pm i(\eta-{\{|\bar{\alpha}_2|,|\bar{\alpha}^{\prime}_2|\}}_{\rm min})$.

4) In regime IV, where $\{|\alpha^{\prime}_1|,|\bar{\alpha}^{\prime}_1|\}\geq \eta$ and $\{|\bar{\alpha}_2|,|\bar{\alpha}^{\prime}_2|\}< \eta$, as shown in Fig.\ref{fig32:subfig:b}, all $\bar{z}$ -zeros form $4N-2$ real zeros, two extra pairs and two boundary conjugate pairs $\pm i(\eta-|\bar{\alpha}_2|)$, $\pm i(\eta-|\bar{\alpha}^{\prime}_2|)$.

5) In regime V, where $|\alpha_1|\geq \eta$, $\{|\alpha^{\prime}_1|,|\bar{\alpha}_2|,|\bar{\alpha}^{\prime}_2|\}< \eta$ or $\{|\alpha_1|,|\bar{\alpha}_2|,|\bar{\alpha}^{\prime}_2|\}< \eta$, $|\alpha^{\prime}_1|\geq \eta$, as shown in Fig.\ref{fig32:subfig:c}, all $\bar{z}$-zeros form $4N-4$ real zeros, two extra pairs and two boundary conjugate pairs $\pi\pm i(\eta-{\{|{\alpha}_1|,|{\alpha}^{\prime}_1|\}}_{\rm min})$, $\pm i(\eta-|\bar{\alpha}_2|)$ and $\pm i(\eta-|\bar{\alpha}^{\prime}_2|)$.

6) In regime VI, where $\{|\alpha_1|,|\alpha^{\prime}_1|,|\bar{\alpha}_2|,|\bar{\alpha}^{\prime}_2|\}< \eta$, as shown in Fig.\ref{fig32:subfig:d}, all $\bar{z}$ -zeros form $4N-6$ real zeros, two extra pairs and two boundary conjugate pairs $\pi\pm i(\eta-|{\alpha}_1|)$, $\pi\pm i(\eta-|{\alpha}^{\prime}_1|)$, $\pm i(\eta-|\bar{\alpha}_2|)$, $\pm i(\eta-|\bar{\alpha}^{\prime}_2|)$.

\begin{table}\renewcommand\arraystretch{2.1}\caption{Patterns of zeros distribution at the ground state for the $D_2^{(1)}$ model and the $D_2^{(2)}$ model. Here, the parameters $\chi_{0}$, $\chi_{0^{\prime}}$, $\chi_1$ and $\chi_2$ are four real numbers that tend to infinity in the thermodynamic limit.}

\resizebox{\textwidth}{!}{
\begin{tabular}{|c|c|c|}
\hline &  {\bf Zeros of the $D_2^{(1)}$ model} & {\bf Zeros of the $D_2^{(2)}$ model}\\

\hline \makecell{Real zeros} & \makecell{$\Lambda^{s+}(u)$: $\{\tilde{z}_k\}$\\ $\tilde{\Lambda}^{s-}(u)$: $\{\tilde{z}_j\}$} & \makecell{$\tilde{\Lambda}_s(u)$: $\{\tilde{z}_k\}$\\ $\tilde{\Lambda}_s(u+i\pi)$: $\{\tilde{z}_k+\pi\}$}\\

\hline Boundary pairs & \makecell{$\Lambda^{s+}(u)$: $\pi\pm i(\eta-|{\alpha}_1|)$,\,$\pi\pm i(\eta-|{\alpha}^{\prime}_1|)$,\\ $\pm i(\eta-|\bar{\alpha}_2|)$, $\pm i(\eta-|\bar{\alpha}^{\prime}_2|)$\\ $\Lambda^{s-}(u)$: \,$\pi\pm i(\eta-|{\gamma}_1|)$,\,
$\pi\pm i(\eta-|{\gamma}^{\prime}_1|)$,\\$\pm i(\eta-|\bar{\gamma}_2|)$,\, $\pm i(\eta-|\bar{\gamma}^{\prime}_2|)$} & \makecell{$\tilde{\Lambda}_s(u)$: $\pi\pm i(\eta-|{\alpha}_1|)$,\,$\pi\pm i(\eta-|{\alpha}^{\prime}_1|)$,\\ $\pm i(\eta-|\bar{\alpha}_2|)$, $\pm i(\eta-|\bar{\alpha}^{\prime}_2|)$\\$\tilde{\Lambda}_s(u+i\pi)$: $\pm i(\eta-|{\alpha}_1|)$,\,
$\pm i(\eta-|{\alpha}^{\prime}_1|)$,\\$\pi\pm i(\eta-|\bar{\alpha}_2|)$,\, $\pi\pm i(\eta-|\bar{\alpha}^{\prime}_2|)$}\\
\hline

Extra pairs & \makecell{$\Lambda^{s+}(u)$: $\pm i\chi_0$\\$\Lambda^{s-}(u)$ $\pm i\chi_{0^{\prime}}$} & \makecell{$\tilde{\Lambda}_s(u)$: $\pm i\chi_1$,\, $\pi\pm i\chi_2$\\$\tilde{\Lambda}_s(u+i\pi)$: $\pi\pm i\chi_1$,\, $\pm i\chi_2$} \\
\hline
\end{tabular}}

\label{open-D21-D22-table}
\end{table}

We find that the zero pattern for the $D_2^{(2)}$ model is different from the one for the model related to the untwisted $D_2^{(1)}$ algebra \cite{NPB9842022}. In Tab.\ref{open-D21-D22-table}, we list the differences between the zeros of the $D_2^{(1)}$ model and the $D_2^{(2)}$ model. We observe that the zeros of the $D_2^{(1)}$ model are composed of the zeros of two independent XXZ spin chains. However, the zeros of the $D_2^{(2)}$ model are composed of the zeros of two staggered XXZ spin chains and these zeros must satisfy the relations $\{\bar z_j=\bar z_k+\pi|j,k=1,\ldots,4N+4\}$. The relationship among these zeros is analogous to that in the Izergin--Korepin model associated with the twisted $A_2^{(2)}$ algebra \cite{Lu24arXiv}.

\section{Surface energy}
\label{SE}
We now study the surface energy of the system based on the patterns of the zeros. The surface energy, induced by the generic boundary fields, is defined as $E_b=E_g-E_p$, where $E_g$ is the ground state energy of the Hamiltonian (\ref{H-D22}) and $E_p$ is the ground state energy of the corresponding periodic chain. In the thermodynamic limit, the distribution of $\tilde{z}$-zeros and the inhomogeneity parameters have continuum densities
\bea
\rho(\tilde{z})=\frac{1}{2N(\tilde{z}_{j+1}-\tilde{z}_{j})}, \quad \sigma(\bar{\theta})=\frac{1}{2N(\bar{\theta}_{l+1}-\bar{\theta}_{l})}.
\eea
In regime I, taking the logarithm of the zeros BAEs (\ref{ba-zeros}) and then subtracting $\bar{\theta}_{j+1}$ and $\bar{\theta}_{j}$, by omitting the $O(N^{-1})$ terms in the thermodynamic limit, we readily obtain
\bea
&&4N\int_{-\pi}^{\pi}b_1(u-\bar{z})\rho(\bar{z})d\bar{z}+b_{\frac{\chi_1+\eta}{\eta}}(u)+b_{\frac{\chi_1-\eta}{\eta}}(u)+b_{\frac{\chi_2+\eta}{\eta}}(u+\pi)+b_{\frac{\chi_2-\eta}{\eta}}(u+\pi)
\no\\[8pt]
&&\hspace{-0.4truecm}=2N\int_{-\pi}^{\pi}[b_2(u-\bar{\theta})+b_2(u+\pi-\bar{\theta})]\sigma(\bar{\theta})d\bar{\theta}+b_2(u)+b_2(u+\pi)-b_1(u)\no\\[8pt]
&&-b_1(u+\pi)+b_{\frac{|\alpha_1|}{\eta}}(u+\pi)+b_{\frac{|\bar{\alpha}_2|}{\eta}}(u)+b_{\frac{|\alpha_1^{\prime}|}{\eta}}(u+\pi)+b_{\frac{|\bar{\alpha}_2^{\prime}|}{\eta}}(u),\label{int-p1}
\eea
where $b_n(u)=\frac{1}{2\pi}\partial_u[\ln\sin(\frac{u}{2}-\frac{in\eta}{2})+\ln\sin(\frac{u}{2}+\frac{in\eta}{2})]=\frac{1}{2\pi}\frac{\sin(u)}{\cosh(n\eta)-\cos u}$. Solving the Eq.(\ref{int-p1}) by Fourier transformation, we have
\bea\label{rho1}
\tilde{\rho}(k)=&&\hspace{-0.6truecm}2N[\tilde{b}_2(k)(1+e^{-i\pi k})]\tilde{\sigma}(k)+\tilde{b}_2(k)(1+e^{-i\pi k})-\tilde{b}_1(k)(1+e^{-i\pi k})\no\\[8pt]
&&\hspace{-0.6truecm}
-\tilde{b}_{\frac{\chi_1+\eta}{\eta}}(k)-\tilde{b}_{\frac{\chi_1-\eta}{\eta}}(k)-(\tilde{b}_{\frac{\chi_2+\eta}{\eta}}(k)+\tilde{b}_{\frac{\chi_2-\eta}{\eta}})e^{-i\pi k}+(\tilde{b}_{\frac{|\alpha_1|}{\eta}}(k)+\tilde{b}_{\frac{|\alpha_1^{\prime}|}{\eta}}(k))e^{-i\pi k}
\no\\[8pt]
&&\hspace{-0.6truecm}+\tilde{b}_{\frac{|\bar{\alpha}_2|}{\eta}}(k)+\tilde{b}_{\frac{|\bar{\alpha}_2^{\prime}|}{\eta}}(k)] /[4N\tilde{b}_1(k)],
\eea
where $\tilde{a}_n(k)$ and $\tilde{b}_n(k)$ are calculated as
\bea
&&\tilde{a}_n(k)=\int_{-\pi}^{\pi}a_n(u)e^{-iuk}du=e^{-|n\eta k|},\\
&&\tilde{b}_n(k)=\int_{-\pi}^{\pi}b_n(u)e^{-iuk}du= sign(k) i e^{-|n\eta k|},
\eea
and the Fourier spectrum $k$ takes integer values. In the homogeneous limit where $\sigma(\bar{\theta})=\delta(\bar{\theta})$ and $\tilde{\sigma}(k)=1$, the inverse Fourier transformation of $\tilde{\rho}(k)$ (\ref{rho1}) is given by
\bea\label{rho11}
\hspace{-0.75truecm}\rho(u)\hspace{-0.08truecm}=&&\hspace{-0.68truecm}\frac{a_1(u)\hspace{-0.08truecm}+\hspace{-0.08truecm}a_1(u\hspace{-0.08truecm}+\hspace{-0.08truecm}\pi)}{2}\hspace{-0.08truecm}+\hspace{-0.08truecm}\frac{1}{4N}\Big[a_1(u)\hspace{-0.08truecm}+\hspace{-0.08truecm}a_1(u\hspace{-0.08truecm}+\hspace{-0.08truecm}\pi)\hspace{-0.08truecm}-\hspace{-0.06truecm}\delta(u)-\delta(u\hspace{-0.08truecm}+\hspace{-0.08truecm}\pi)\hspace{-0.08truecm}-\hspace{-0.08truecm}a_{\frac{\chi_1}{\eta}}(u)\hspace{-0.08truecm}-\hspace{-0.08truecm}a_{\frac{\chi_1-2\eta}{\eta}}(u)\no\\[8pt]
\hspace{-0.85truecm}&&\hspace{-0.68truecm}-a_{\frac{\chi_2}{\eta}}(u\hspace{-0.08truecm}+\hspace{-0.08truecm}\pi)\hspace{-0.08truecm}-\hspace{-0.08truecm}a_{\frac{\chi_1-2\eta}{\eta}}(u\hspace{-0.08truecm}+\hspace{-0.08truecm}\pi)\hspace{-0.08truecm}+\hspace{-0.08truecm}a_{\frac{|\alpha_1|-\eta}{\eta}}(u\hspace{-0.08truecm}+\hspace{-0.08truecm}\pi)\hspace{-0.08truecm}+\hspace{-0.08truecm}a_{\frac{|\alpha_1^{\prime}|-\eta}{\eta}}(u\hspace{-0.08truecm}+\hspace{-0.08truecm}\pi)\hspace{-0.08truecm}+\hspace{-0.08truecm}a_{\frac{|\bar{\alpha}_2|-\eta}{\eta}}(u)\hspace{-0.08truecm}+\hspace{-0.08truecm}a_{\frac{|\bar{\alpha}_2^{\prime}|-\eta}{\eta}}(u)\Big],
\eea
where $\delta(u)$ is the Dirac delta function. In the thermodynamic limit, the ground state energy of the Hamiltonian (\ref{H-D22}) in regime I can be expressed as
\bea\label{Eg1}
E_{g1}=&&\hspace{-0.6truecm}-4\pi N\int_{-\pi}^{\pi}[a_1(z)+a_1(z+\pi)]\rho(z)dz-\tanh(2\eta)\no\\[8pt]
&&\hspace{-0.4truecm}-2\pi\Big[a_1(i\chi_1)+a_1(i\chi_1+\pi)+a_1(i\chi_2)+a_1(i\chi_2+\pi)\Big],
\eea
The ground state energy of the system with periodic boundary conditions can be obtained similarly
\bea
E_p=&&\hspace{-0.6truecm}-2\pi N\int_{-\pi}^{\pi}[a_1(z)+a_1(z+\pi)]^2d z-\tanh(2\eta)\no\\[6pt]=&&\hspace{-0.6truecm}-2N(\coth \eta+\tanh \eta)-\tanh(2\eta).
\eea

After tedious calculations and considering $\chi_1,\chi_2\to \infty$ in the thermodynamic limit, we obtain the surface energy $E_{b1}$ in regime I as
\bea
&&\hspace{-1truecm}E_{b1}=e_b(\alpha_1,\alpha_2)+e_b(\alpha_1^{\prime},\alpha_2^{\prime})+e_{b0},\label{Eb1}\\[8pt]
&&\hspace{-1truecm}e_b(x,y)=-\frac{1}{2}(\coth \frac{|x|}{2}+\tanh \frac{|x|}{2}+\coth \frac{|y|}{2}+\tanh \frac{|y|}{2}),\\[6pt]
&&\hspace{-1truecm}e_{b0}=\coth \frac{\eta}{2}+\tanh \frac{\eta}{2}-\coth \eta-\tanh \eta+4.
\eea
where $e_b(x,y)$ indicates the contribution of one boundary field and $e_{b0}$ is the surface energy induced by the free open boundary. For the regime II, using the similar procedure to that applied in regime I, we have
\bea
&&\hspace{-1truecm}E_{b2}=E_{b1}+e_{b2}(\{|\bar{\alpha}_2|,|\bar{\alpha}^{\prime}_2|\}_{\rm min}),\no
\\[6pt]&&\hspace{-1truecm}e_{b2}(x)=\frac{1}{2}(\coth \frac{|x|}{2}+\tanh \frac{|x|}{2}+\coth \frac{2\eta-|x|}{2}+\tanh \frac{2\eta-|x|}{2})\no\\[6pt]&&\quad-2\pi\Big[a_1(i\eta-i|x|)+a_1(i\eta -i|x|+\pi) \Big].
\eea
One can find that the bare contributions of the boundary pairs to the ground-state energy are exactly canceled by those of the back flow of continuum root density, namely
\bea
e_{b2}(x)=0,\qquad 0<|x|<\eta.
\eea
A simple proof is as follows:
\bea
&&2\pi[a_1(i\eta-i|x|)+a_1(i\eta-i|x|+\pi)]\no\\[6pt]
=&&\hspace{-0.6truecm}\frac{1}{2i}\Big[\cot(\frac{i\eta-i|x|}{2}-\frac{i\eta}{2})-\cot(\frac{i\eta-i|x|}{2}+\frac{i\eta}{2})\no\\[6pt]&&+\cot(\frac{i\eta-i|x|+\pi}{2}-\frac{i\eta}{2})-\cot(\frac{i\eta-i|x|+\pi}{2}+\frac{i\eta}{2})\Big]\no\\[6pt]
=&&\hspace{-0.6truecm}\frac{1}{2}[\coth(\frac{|x|}{2})+\coth(\frac{2\eta-|x|}{2})+\tanh(\frac{|x|}{2})+\tanh(\frac{2\eta-|x|}{2})].\label{Pat}
\eea
Similar results are also observed in regimes III-VI. Therefore, the surface energies in all regimes can be expressed in the form of Eq.(\ref{Eb1}).
The surface energy with the boundary parameter $r={\rm Re}(s_1)$ is shown in Fig.\ref{fig-surfaceE} below. The analytic results obtained from (\ref{Eb1}) and the numerical ones calculated from DMRG are consistent with each other.

\begin{figure}[htp]
\centering
\includegraphics[width=8cm]{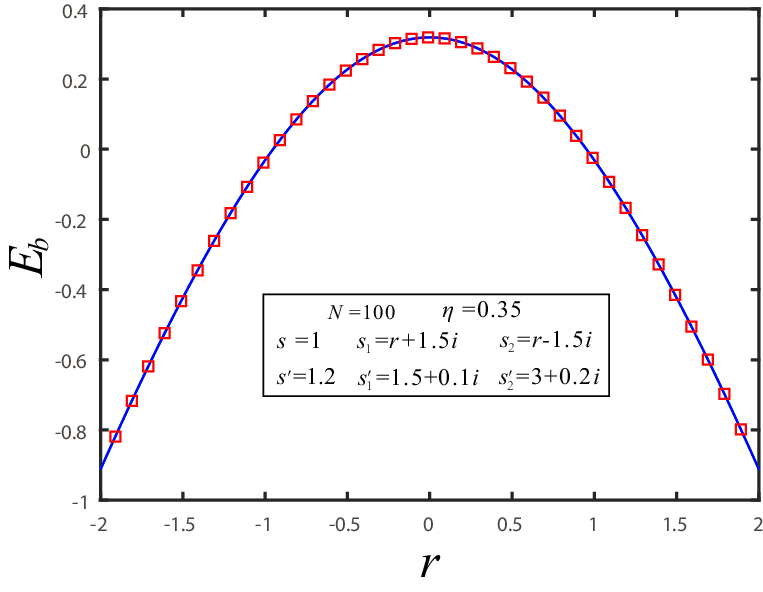}
  \caption{Surface energy $E_b$ versus the boundary parameter $r={\rm Re}(s_1)$. The line indicates the analytic results and the squares indicate the DMRG results for $N=100$. They are consistent with each other very well.}
\label{fig-surfaceE}
\end{figure}

\section{Elementary excitations}
\label{excitations} 
\begin{figure}[htp]
\centering
\subfigure{\label{5a}
\includegraphics[scale=0.62]{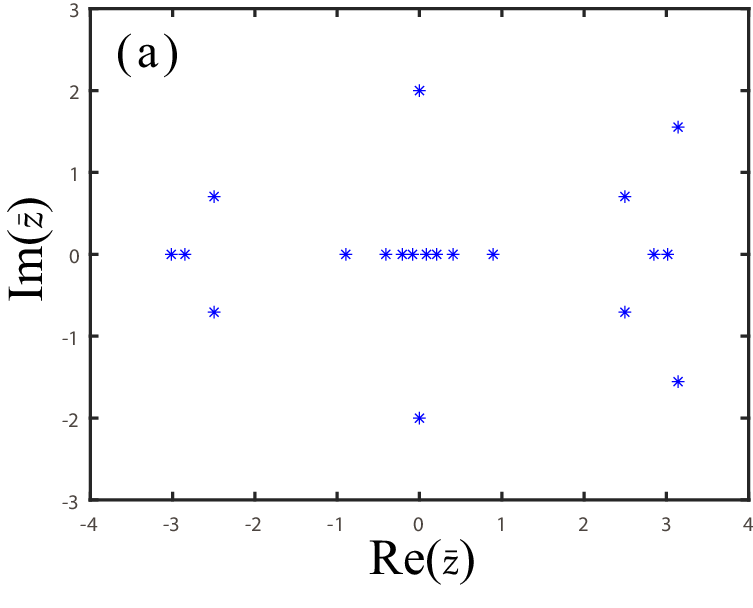}
}
\subfigure{\label{5b}
\includegraphics[scale=0.62]{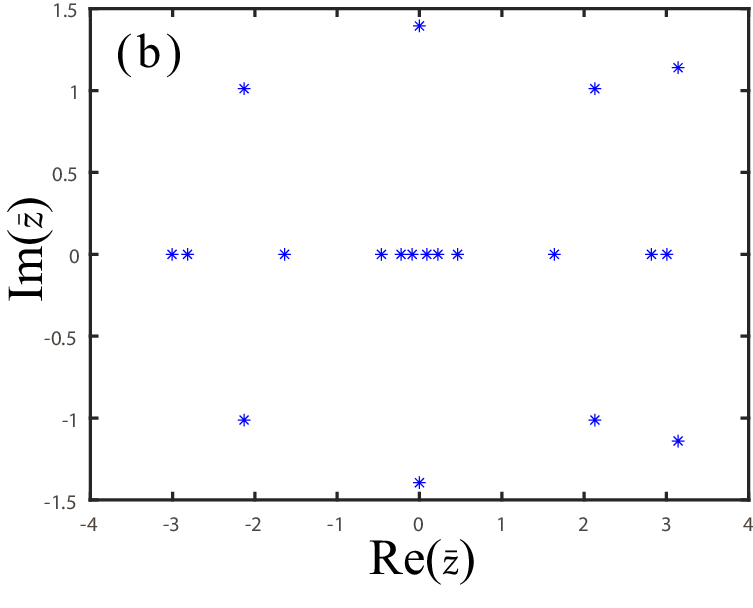}
}
\caption{The distribution of $\bar{z}$-zeros for $\{\bar{\theta}_j= 0|j=1,\cdots,2N\}$ at the excited states. (a) The 19th excited state with $n=2$. (b) The 39th excited state with $n=3$.}\label{fig5}
\end{figure}

\setcounter{equation}{0}
In this section we study the bulk elementary excitations and boundary excitations in the system. We first consider the elementary excitations in the bulk. The bulk excitations in different regimes of boundary parameters are the same. From
the patterns of the zeros in the low-lying excited states, we find that the excitations can be
characterized by putting four real zeros in the two conjugate pairs $\{\tilde{z}_n\pm in\eta,\,-\tilde{z}_n\pm in\eta|n\geq 2\}$.
In order to see the bulk excitations more clearly,
we show the patterns of the zeros at the $n=2$ and $n=3$ excited states in Fig. \ref{fig5}, where the ground state is still in regime I.

\begin{figure}[htp]
\centering
\subfigure{
\includegraphics[scale=0.8]{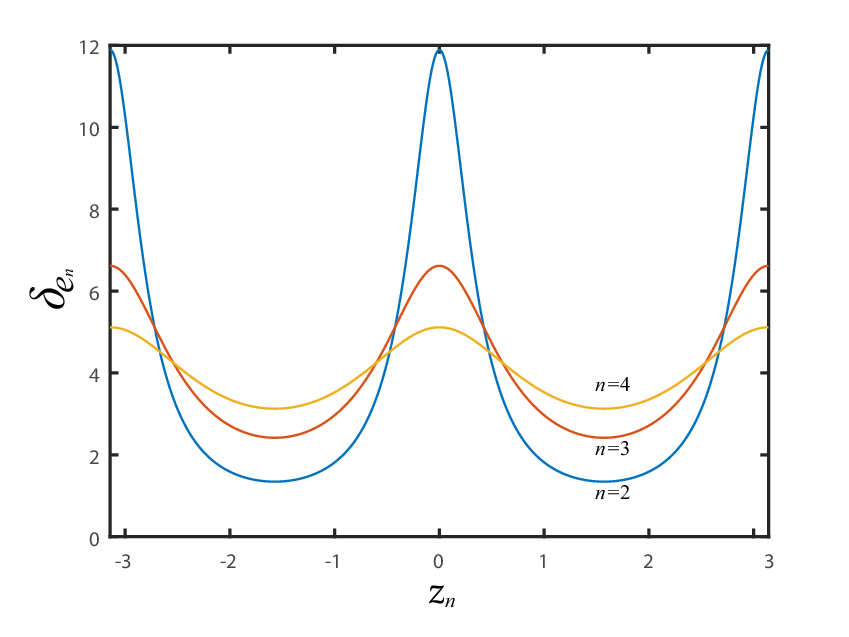}
}
\caption{The excited energies $\delta_{e_n}$ versus $\tilde{z}_n$ in the thermodynamic limit..}\label{fig6}
\end{figure}

In the thermodynamic limit, the density
difference $\delta\tilde{\rho}_{e_n}(k)$ between the ground state and the excited state is
\bea
\delta{\rho}_{e_n}(u)=-\frac{1}{4N}(a_n(u+\tilde{z}_n)+a_n(u-\tilde{z}_n)+a_{n-2}(u+\tilde{z}_n)+a_{n-2}(u-\tilde{z}_n)),
\eea
where $a_n(u)=\frac{1}{2\pi}\frac{\sinh(n\eta)}{\cosh(n\eta)-\cos u}$ and $\tilde{z}_n\in(-\pi,\pi]$. The related elementary excitation energy is
\bea
\delta_{e_n}=4\pi [a_{n-1}(\tilde{z}_n)+a_{n-1}(\tilde{z}_n+\pi)].
\eea
The excited energies $\delta_{e_n}$ versus $\tilde{z}_n$ are shown in Fig. \ref{fig6}.

Now, we focus on the boundary excitations. By comparing the distributions of the zeros for the ground state and the excited states, we find that the boundary excitations can exist in regimes II-VI of boundary parameters. The
typical boundary excitation is putting the boundary pairs from $\pi\pm i(\eta-|x|)$ to $\pi\pm i(\eta+|x|)$ or from $\pm i(\eta-|y|)$ to $\pm i(\eta+|y)$, where $x=\alpha_1$ or $\alpha_1^{\prime}$ and $y=\bar{\alpha}_2$ or $\bar{\alpha}_2^{\prime}$. In the excitation process, one extra pair at origin or at ${\rm Re}\{\bar{z}_j\}=\pi$ of the ground state jumps to real axis.

\begin{figure}[htp]
\centering
\subfigure{\label{7a}
\includegraphics[scale=0.6]{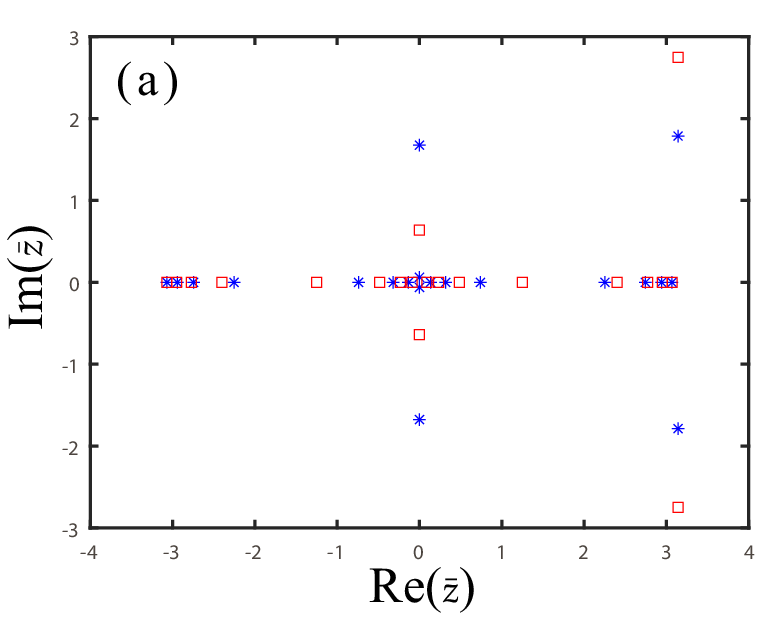}
}
\subfigure{\label{7b}
\includegraphics[scale=0.6]{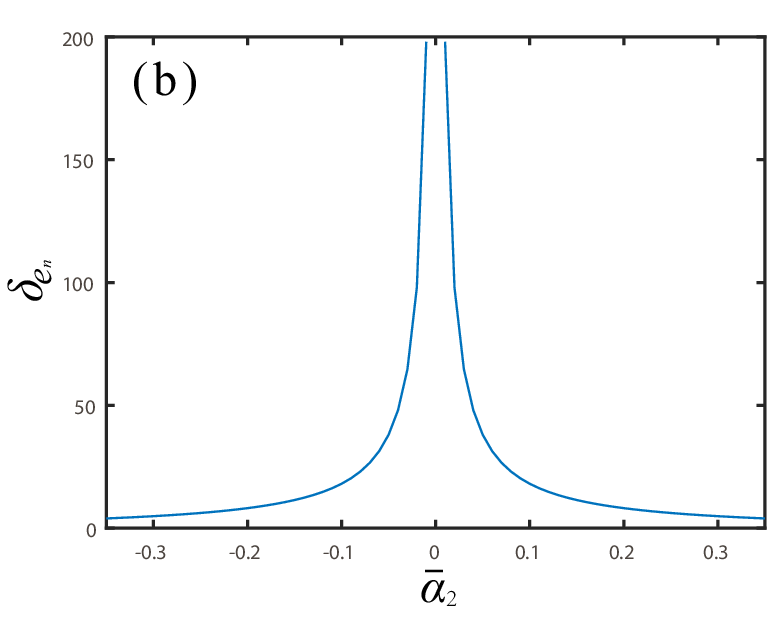}
}
\caption{(a) The distribution of $\bar{z}$-zeros for $\{\bar{\theta}_j= 0|j=1,\cdots,2N\}$   with $N=4$, $\eta=0.35$, $s=1$, $s_1=4$, $s_2=4$, $s'=0.6$, $s'_1=1$ and $s'_2=2.014$. Here the blue asterisks represent the pattern of zeros at the ground state and the red squares denote those at the 19th excited state with boundary pairs $\pm i(\eta-|\bar{\alpha}_2|)$, $\pm i(\eta+|\bar{\alpha}_2|)$.
(b) The boundary excitation energy versus $\bar{\alpha}_2$ in the thermodynamic limit.}\label{fig7}
\end{figure}

As an example, we show in Fig.\ref{7a} the patterns of the zeros for the ground state (blue asterisks) and the 19th excited state (red squares) with boundary pairs $\pm i(\eta+|\bar{\alpha}_2|)$ in regime II with $N=4$. In the thermodynamic limit, the resulting density change between ground and excited states reads
\bea
\delta{\rho}_{\bar{\alpha}_2}(u)=\frac{1}{4N}[a_{\frac{\eta-|\bar{\alpha}_2|}{\eta}}(u)-a_{\frac{\eta+|\bar{\alpha}_2|}{\eta}}(u)+a_{\frac{\chi_1}{\eta}}(u)+a_{\frac{\chi_1-2\eta}{\eta}}(u)].
\eea
The related excited energy is
\bea
\delta_{e}(|\bar{\alpha}_2|)=\coth \frac{|\bar{\alpha}_2|}{2}+\tanh \frac{|\bar{\alpha}_2|}{2}-2.
\eea
The excited energies versus $\bar{\alpha}_2$ are shown in Fig.\ref{7b}. From similar
idea and after tedious calculations, we obtain the boundary excitations in the regimes II-VI, as listed in Tab.\ref{Boundary-excitations-table}.
\begin{table}\centering\renewcommand\arraystretch{1.2}\caption{Boundary excitations in the regimes II-VI}

\resizebox{0.5\textwidth}{!}{
\begin{tabular}{|c|c|}
\hline &  {\bf Boundary excitations} \\

\hline $\mathrm{II}$ & $\delta_{e}(\{|\bar{\alpha}_2|,|\bar{\alpha}_2^{\prime}|\}_{\rm min})$\\

\hline $\mathrm{III}$ & $\makecell{\delta_{e}(\{|{\alpha}_1|,|{\alpha}_1^{\prime}|\}_{\rm min}),\hspace{4pt} \delta_{e}(\{|\bar{\alpha}_2|,|\bar{\alpha}_2^{\prime}|\}_{\rm min}),\\
\delta_{e}(\{|{\alpha}_1|,|{\alpha}_1^{\prime}|\}_{\rm min})+\delta_{e}(\{|\bar{\alpha}_2|,|\bar{\alpha}_2^{\prime}|\}_{\rm min}) }$\\

\hline $\mathrm{IV}$ & $\makecell{\delta_{e}(|\bar{\alpha}_2|),\hspace{4pt} \delta_{e}(|\bar{\alpha}^{\prime}_2|),\\ 2+\delta_{e}(|\bar{\alpha}_2|)+\delta_{e}(|\bar{\alpha}^{\prime}_2|)}$ \\
\hline

$\mathrm{V}$ & $\makecell{\delta_{e}(\{|{\alpha}_1|,|{\alpha}_1^{\prime}|\}_{\rm min}),\hspace{4pt}\delta_{e}(|\bar{\alpha}_2|),\hspace{4pt} \delta_{e}(|\bar{\alpha}^{\prime}_2|),\\ 2+\delta_{e}(|\bar{\alpha}_2|)+\delta_{e}(|\bar{\alpha}^{\prime}_2|)}$\\

\hline
$\mathrm{VI}$ & $\makecell{\delta_{e}(|{\alpha}_1|),\hspace{4pt}\delta_{e}(|{\alpha}^{\prime}_1|),\hspace{4pt}\delta_{e}(|\bar{\alpha}_2|),\hspace{4pt} \delta_{e}(|\bar{\alpha}^{\prime}_2|),\\ 2+\delta_{e}(|{\alpha}_1|)+\delta_{e}(|{\alpha}^{\prime}_1|),\hspace{4pt} 2+\delta_{e}(|\bar{\alpha}_2|)+\delta_{e}(|\bar{\alpha}^{\prime}_2|)}$\\
\hline
\end{tabular}}

\label{Boundary-excitations-table}
\end{table}

\newpage
\section{Conclusions}
\label{Con}
In this paper, we have generalized the $t$-$W$ method to the $D_2^{(2)}$ spin chain with generic integrable boundaries. With the help of factorization relation (\ref{Factor-t}) of the transfer matrix, the eigenvalue problem of the $D_2^{(2)}$ spin chain is transformed into the eigenvalue problem of the staggered XXZ spin chain. By parametrizing the eigenvalues of the transfer matrix for the staggered XXZ spin chain in terms of its zeros, we derive a set of homogeneous BAEs for the zeros of the transfer matrix eigenvalues. The zero distribution patterns in different regimes of the boundary parameters are obtained by solving the BAEs.
Due to the symmetry of the twisted quantum affine algebra,
Unlike the model related to the untwisted $D_2^{(1)}$ algebra, the zeros of two staggered XXZ spin chains must satisfy certain constraint for them to provide the correct zeros of the $D_2^{(2)}$ model.
Based on the patterns of the zeros, we have derived the densities of the zeros, exact surface energies and elementary excitations in different regimes of the boundary parameters. The results of the surface energy indicate that the boundary fields on both sides of the system are independent of each other in the thermodynamic limit. Indeed,
the unparallel boundary magnetic fields will induce the helical states, which can be calculated with the help of the obtained eigenvalues. Through these helical states, persistent currents that exist within the system can be further studied.

Interesting open problems for future work include the construction of the $t$-$W$ relation for the quantum transfer matrix of the system  at finite temperature and the calculation of the corresponding physical quantities such as free energy and specific heat. It is also important to explore the quench dynamics of the system through the combination of the boundary quantum transfer matrix method \cite{JSM2017023106,
JSM2018113102} and the $t$-$W$ method.
The method and process presented in this paper can also be extended to other models with twisted affine algebra symmetries \cite{DGZ96}.

\section*{Acknowledgments}

We acknowledge the financial support from Australian Research Council Discovery Project DP190101529, Future Fellowship FT180100099, National Key R$\&$D Program of China (Grant No.2021YFA1402104),
China Postdoctoral Science Foundation Fellowship 2020M680724, National Natural Science Foundation of China (Grant Nos. 12434006, 12247103 and 12247179), the Major Basic Research Program of Natural Science of Shaanxi Province (Grant Nos. 2021JCW-19 and 2017ZDJC-32), and the Strategic Priority Research Program of the Chinese Academy of Sciences (Grant No. XDB33000000).

\end{document}